# High-throughput study of the phase constitution of the thin film system Mg–Mn–Al–O in relation to Li recovery from slags

Florian Lourens[a], Ellen Suhr[a], Alena Schnickmann[b], Thomas Schirmer[b], Alfred Ludwig[a, *]

[a] *Materials Discovery and Interfaces, Institute for Materials, Ruhr University Bochum, Universitätsstrasse 150, 44780 Bochum, Germany*

[b] *Department of Mineralogy, Geochemistry, Salt Deposits, Institute of Disposal Research, Clausthal University of Technology, Adolph-Roemer-Str. 2A, 38678 Clausthal-Zellerfeld, Germany*

*\* Corresponding author: alfred.ludwig@rub.de*

Keywords: Magnetron sputtering, combinatorial materials science, engineered artificial minerals

**Abstract**

The increasing importance of recycling makes the recovery of valuable elements from slags interesting, e.g., by the concept of engineered artificial minerals (EnAMs). In this concept, it is aimed for the formation of EnAMs, meaning phase(s) with a high content of the to-be-recovered element(s) from slags of pyrometallurgical recycling processes. For this, understanding the phase constitution of the slag systems is of high importance. The system Mg–Mn–Al–O is a metal oxide slag subsystem from Li-ion battery recycling, that is critical for the formation of spinel phases, which are competing phases to the possible Li-containing EnAM phase $LiAlO_2$. Here, the phase constitution was investigated using a thin film materials library that covers the composition space $(Mg_{14–69}Mn_{11–38}Al_{14–74})O_x$. By means of high-throughput energy-dispersive X-ray spectroscopy and X-ray diffraction, the formation of the spinel solid solution phase was confirmed for a wide composition space. Increasing preferential orientation of the spinel solid solution along (400) with increasing Mg content was identified. X-ray photoelectron spectroscopy was used to measure the near-surface composition of selected areas of the materials library, and detailed peak fitting of the Mn 2p3/2 region revealed the Mn oxidation state to be a mixture of $Mn^{2+}$ and $Mn^{3+}$. For one measurement area of the materials library containing equal atomic amounts of Mg, Mn and Al, transmission electron microscopy showed that the approximately 420 nm-thick film consists of columnar spinel grains with Mg, Mn and Al being evenly distributed. Based on these results, we suggest that the shown high likelihood of spinel formation in slags might be influenced by controlling the Mn oxidation state to enable the formation of desirable EnAM phases.

**Introduction**

Pyrometallurgical processes are common for the recycling of various wastes such as spent batteries [1–4], circuit boards [5] and catalysts [2], as they are easily scalable and suitable for many different raw materials. In pyrometallurgical processes though, besides the alloy phase containing the valuable metals targeted to be recovered (e.g. Co, Ni, Cu), there is the slag phase, where less noble and yet sometimes technologically relevant elements (e.g. Li, Al, Mg, Mn) are forming complex compounds [1,2]. The recovery of valuable slag elements such as Li [6] requires hydrometallurgical processes, that are cost-intensive, time-consuming, and sometimes environmentally harmful. Therefore, slags are frequently directly used in applications with minor material requirements, e.g. as additives in cement manufacturing, in order to waive any further slag processing [3,6]. This squandering of slag elements contradicts sustainability and resource-saving premises.

Due to its high $O_2$ affinity, Li is an element found in the slag from pyrometallurgical recycling processes of Li-ion batteries (LIB), along with Al, Mn, Mg, Si, and Ca. The variety of Li-containing compounds in slags is large (oxides, silicates / crystalline, amorphous) [7,8]. For economic Li recovery, it would be advantageous if Li could be enriched in an easily separable compound that meets the following criteria:

a)      High enrichment factor for Li

b)      No competing Li-containing early crystallizates

c)      Favorable processing properties (crystal habitus / morphology, surface / magnetic properties, …)

For the formation of such a Li-containing compound, the slag must be modified and/or be post-treated before cooling in a separate reactor. This is the idea behind the Engineered Artificial Mineral (EnAM) approach [9]. A promising EnAM candidate phase for Li recovery is $LiAlO_2$, as it is an early crystallizate containing a large amount of Li and can be separated from the rest of the slag by flotation [10]. However, competing early crystallizates exist: refractory spinel solid solutions, that gather Al from the melt could completely suppress $LiAlO_2$ formation [9]. Besides Al, Mg and Mn are typical spinel-forming elements. Mn is increasingly used in Mg–Mn–Al–O-based cathode materials as a low-cost and non-toxic alternative to Co- or Ni-based materials [4], leading to considerable Mn contents in the slag. This challenge of competing crystallization of spinel phases and $LiAlO_2$ can either be overcome by suppressing the spinel formation, or by replacing $LiAlO_2$ with another EnAM.

In this work, the phase constitution of the Li-slag subsystem Mg–Mn–Al–O is investigated using a thin film materials library and high-throughput characterization. This system comprises the elements typically involved in the spinel formation in slag phases. One idea to influence the spinel formation is to avoid the more reduced $Mn^{2+}$ and $Mn^{3+}$ species in favor of $Mn^{4+}$. While $Mn^{2+/3+}$ occupy the tetrahedral and octahedral sites in the spinel lattice, the hypothesis is that $Mn^{4+}$ cannot form oxides with Al or Mg (at least no reference was found in literature). As a result, $Mn^{4+}$ would be available in the slag to form the Li manganate $Li^{1+}_2Mn^{4+}O^{2-}_3$, which could then be considered a new EnAM. In addition, by preventing Mn from being part of the spinel formation, spinel formation may be reduced or even prevented altogether, possibly making $LiAlO_2$ formation more likely. In this case, both $Li_2MnO_3$ and $LiAlO_2$ can be considered as EnAMs.

The known compounds and solid solutions in the subsystems of Mg-Mn-Al-O are typically spinel phases. In the systems $MnO-Al_2O_3$ and $MgO-Al_2O_3$, there are the cubic spinels $MnAl_2O_4$ (galaxite) [11] and $MgAl_2O_4$ [12]. They are known to form a solid solution by substituting $Mg^{2+}$ and $Mn^{2+}$ [13]. A mixture of $Mn^{2+}$ and $Mn^{3+}$ ($MnO-Mn_2O_3$ system) can form the spinel $Mn^{2+}Mn^{3+}_2O_4$ (hausmannite), which below 1172°C forms a distorted tetragonal spinel due the Jahn-Teller-active cation $Mn^{3+}$ [11,14,15]. In its cubic form, the octahedral sites of hausmannite are occupied by a mixture of $Mn^{2+}$, $Mn^{3+}$ and $Mn^{4+}$ [16]. Another $Mn^{3+}$-containing spinel forming a tetragonal lattice due to the Jahn-Teller deformation is $MgMn_2O_4$ ($MgO-Mn_2O_3$ system) [17–20]. $MgMn_2O_4$ and hausmannite can form tetragonal or cubic spinel solid solutions, containing mixtures of $Mn^{2+}$ and $Mn^{3+}$, and $Mn^{2+}$, $Mn^{3+}$ and $Mn^{4+}$, respectively [21]. Theoretically, solid solutions of hausmannite or $MgMn_2O_4$ with the cubic spinels $Mg/MnAl_2O_4$ are only possible when the Jahn-Teller deformation is offset, e.g. at high temperatures. Beyond that, $Al^{3+}$ can be incorporated into $MgMn_2O_4$, which by substituting $Mn^{3+}$ causes a structural change from tetragonal to cubic, making $MgMn_2O_4$-$MgAl_2O_4$ solid solutions possible [19]. Both MgO and MnO have NaCl-type crystal structures. Their quasi-binary phase diagram shows a miscibility gap (two phase region MgO + MnO) at temperatures below 500°C [21].

**Methods**

A Mg–Mn–Al–O thin film materials library (ML) was fabricated by reactive magnetron co-sputtering. The used sputter system (Creavac, Germany) has a sputter-up configuration with three confocal cathodes that are positioned on a horizontal circle with circular arcs of 120° between them. The sputter targets were elemental Mg (50.8 mm diameter, 99.95 % purity, AJA

International), Mn (101.6 mm diameter, 99.95 % purity, Sindlhauser Materials) and Al (50.8 mm diameter, 100 % purity, Kurt J. Lesker Company). Their relative locations are schematically visualized in Fig. 1. The base pressure prior to the deposition was 9.8 x10$^{-6}$ Pa. A Si wafer (100 mm diameter) with native oxide was used as the substrate, which was heated to 300°C, with a pre-heating period of 120 min. The deposition pressure was 2 Pa. The gas flows of Ar (6.0, Praxair) and O$_2$ (6.0, Praxair) were 40 SCCM and 2 SCCM, respectively. During the deposition duration of 4 h, 90 W (DC) was applied on the Mg target, 120 W (RF) was applied on the Mn target and 70 W (DC) was applied on the Al target.

An array of 342 measurement areas (MAs) of 4.5 x 4.5 mm$^2$ each were defined across the ML (Fig. 1). All MAs were investigated by high-throughput energy dispersive X-ray spectroscopy (EDX) and X-ray diffraction (XRD) mappings. Additionally, 37 selected MAs were measured by X-ray photoelectron spectroscopy (XPS) (marked with X in Fig. 1). One MA (marked with T in Fig. 1) was used to prepare a thin cross-sectional sample ("lamella") by focused ion beam (FIB) to be investigated by transmission electron microscopy (TEM).

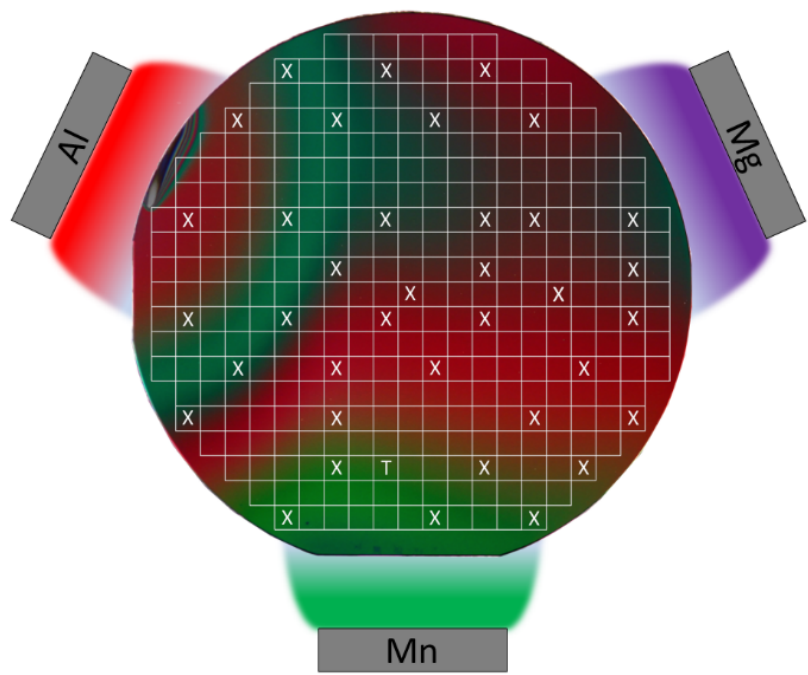

Figure 1: Photograph of the Mg–Mn–Al–O thin film materials library (100 mm diameter), with an overlayed schematic visualization of the grid of 342 MAs, all of which were investigated by EDX and XRD. The MAs marked with X were additionally measured by XPS, and the one marked with T by TEM. The locations of the Mg, Mn and Al sputter targets are indicated (not to scale, arbitrary colors).

The EDX mapping was carried out on a JEOL 5800LV scanning electron microscope (SEM) equipped with an Oxford INCA X-act EDX detector. The electron acceleration voltage was 20 kV, the acquisition time per MA was 60 s, the working distance was 10 mm and the magnification was 600x.

The XRD mapping was done on a Bruker D8 Discover with Bragg-Brentano geometry, using Cu Kα radiation collimated to 1 mm with a divergence of < 0.007° and a VANTEC-500 area detector with a sample-to-detector distance of 149 mm. At each MA, three frames at 2θ of 25°, 45° and 65° were recorded stepwise, each covering 2θ increments of ±15°. XRD patterns were obtained by integrating the detector frames. For the phase identification, the measured patterns were compared to reference patterns form the Inorganic Crystal Structure Database (ICSD) [22].

The TEM sample was prepared using a Helios NanoLab G4 CX FIB system, which is a dual beam system with an electron and a Ga-ion beam. It is equipped with a micromanipulator (EasyLift, Oxford Instruments) and gas injection systems that were used for the deposition of protective C and Pt layers. The prepared lamella for TEM had an initial width of approximately 5 μm and a final thickness of < 100 nm. The TEM measurements, including TEM and scanning TEM (STEM), dark field (DF) imaging and STEM EDX measurements, were carried out on a Jeol Neoarm aberration-corrected TEM, operated at 200 kV. The electron diffraction data was analyzed using CrysTBox [23] by comparing the rotational averaged diffraction patterns to the $MgAl_2O_4$ cif file from the ICSD database (ICSD 411021) [22].

XPS measurements were carried out on a Kratos Axis Nova, using a monochromatic Al Kα X-ray source operated at 180 W (15 mA, 12 kV), and a delay-line detector with a pass energy of 20 eV. The electron emission angle was 0°, the charge neutralizer was turned on, and there was no sputter etching applied. The pressure during the measurements was below $3.2 \times 10^{-6}$ Pa. For each of the selected MAs, high-resolution scans of 300 x 700 $\mu m^2$ analysis areas for the Mg 1s, Mn 2p, Al 2p, O 1s and C 1s regions were measured. The latter was only used for charge correction (adventitious C-C at 284.8 eV), while the others were used for quantification, using Shirley backgrounds and the Kratos ESCApe software with its pre-defined relative sensitivity factors.

**Results**

As determined by the high-throughput EDX mapping, the large compositional range of the as-deposited ML comprises the desired metal gradients 4.7 – 24.9 at. % Mg, 3.4 – 12.6 at. % Mn, 5.1 – 24.6 at. % Al, and an almost constant O content (63.9 – 69.9 at. %). The maxima and minima of the metal gradients shown in Figs. 2 a) – c) correspond to the shortest and longest distances to the sputter targets, respectively (see Fig. 1). The measured O content shown in Fig. 2 d) is the collective of the thin film and the native oxide of the Si wafer. To precisely characterize the composition space of the ML without bias from the wafer, excluding O leads to the normalized composition space of $(Mg_{14–69}Mn_{11–38}Al_{14–74})O_X$.

The phase distribution of the as-deposited ML identified by the high-throughput XRD mapping is given in Fig. 3 a), alongside five example patterns in Fig. 3 b). The patterns of 173 MAs (green in Fig. 3 a) indicate cubic spinel solid solution, as they show all or at least most of the expected diffraction peaks of cubic spinel with varying degrees of deviation from the reference diffraction angles of the possible stoichiometric spinels $MgMn_2O_4$, $MnAl_2O_4$ and $MgAl_2O_4$ (all space group number 227). The reference diffraction angles of the three spinel phases are included in Fig. 3 b). There is a general trend for the spinel peaks to shift to lower diffraction angles the further the MAs are located on the right of the ML, which is accompanied by the disappearance of the peaks except for the (400) one in the 2θ = 42 – 45° range. This leads to 99 MAs (red in Fig. 3 a) only showing the (400) spinel peak, still shifting to lower 2θ the further the MAs are on the right of the ML. These 99 MAs can therefore be characterized as spinel solid solution, with all grains being oriented along their (400) planes parallel to the wafer surface. Further, there are 18 Mg-rich MAs (blue in Fig. 3 a) identified to have MgO structure (cubic NaCl-type structure, space group number 225). In Fig 3 b), it can be observed that the patterns of spinel (400) (red) and MgO (blue) look quite similar. Nonetheless, MgO is identified by the peak at 42.9° and the significantly weaker peak at 36.9°, that in combination do not agree with spinel, but accurately with (200) and (111) MgO, respectively. It is possible though, that there are MAs where the coexistence of MgO and spinel was overlooked due to the similar diffractograms. In fact, this is quite likely, because MgO cannot form a solid solution with spinel, so between the "Spinel (400)" and "MgO" MAs in Fig. 3 a), there should be regions where both phases coexist. In general, the MgO (200) peaks are noticeably broader than the spinel (400) ones, indicating smaller grain sizes for MgO compared to spinel. Besides the crystalline MAs, there are 52 X-ray amorphous MAs (black in Fig. 3 a), which are predominantly Al-rich MAs located on the left of the ML, besides of five Mg-rich MAs on the top right of the ML. The phase distribution of the normalized Mg-Mn-Al compositions is visualized in Fig. 4 in a ternary triangle. The MgO – spinel (400) and spinel – amorphous boundaries appear at approximately constant Mg contents of 63 and 22 at. %, respectively. The boundary between spinel and spinel (400) also appears roughly at a straight line, namely the intersect of 53 at. % Mg and 68 at. % Al.

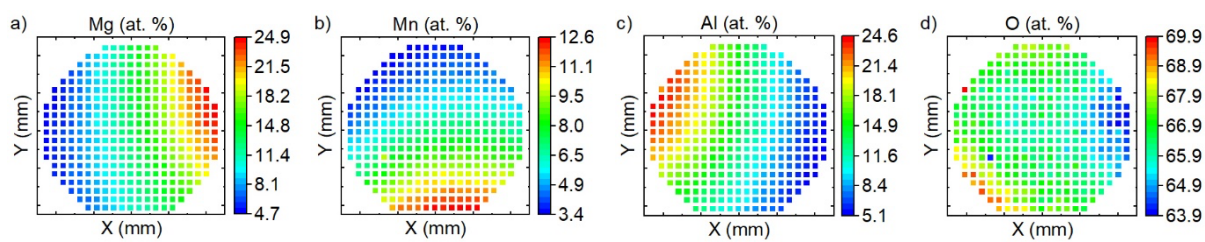

Figure 2: Compositional gradients of the as-deposited thin film materials library: Color-coded elemental contents of a) Mg, b) Mn, c) Al and d) O, determined using high-throughput EDX of the 342 MAs.

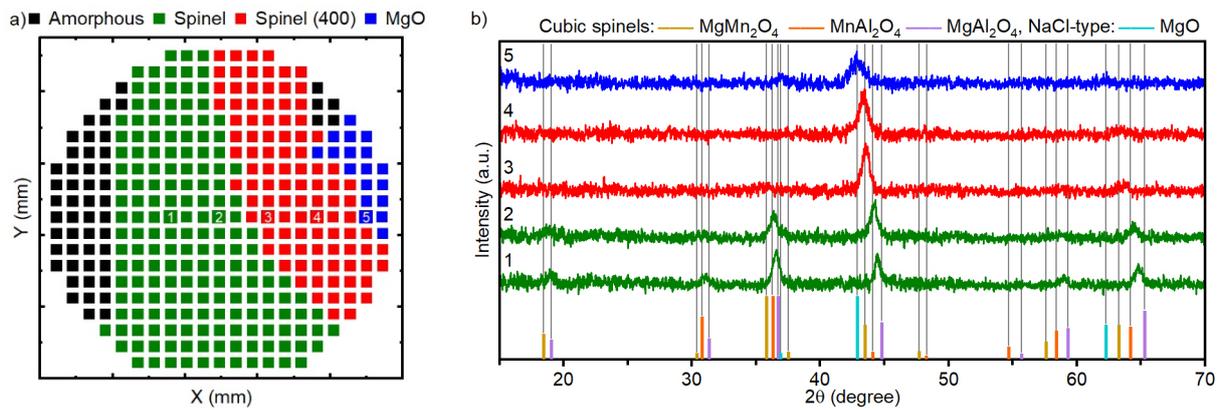

Figure 3: a) Color-coded visualization of the phase distribution across the 342 MAs of the ML. b) Stacked example XRD patterns (intensity in dependance of diffraction angle 2θ) of the MAs one to five of a). In b), the reference peak positions and relative intensities of MgMn$_2$O$_4$ (ICSD 34148), MnAl$_2$O$_4$ (ICSD 252228), MgAl$_2$O$_4$ (ICSD 411021) and MgO (ICSD 61325) are included as vertical lines on the X-axis.

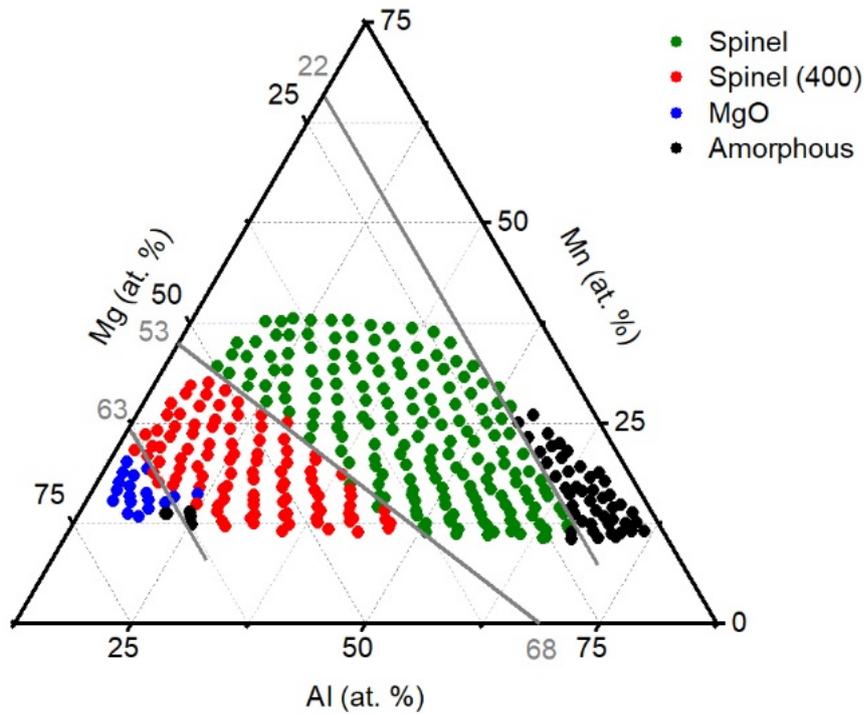

Figure 4: Visualization of the phase distribution in dependance of the normalized Mg–Mn–Al compositions. The grey lines indicate constant Mg contents of 22 at. % and 63 at. %, and the intersect of 53 at. % Mg and 68 at. % Al.

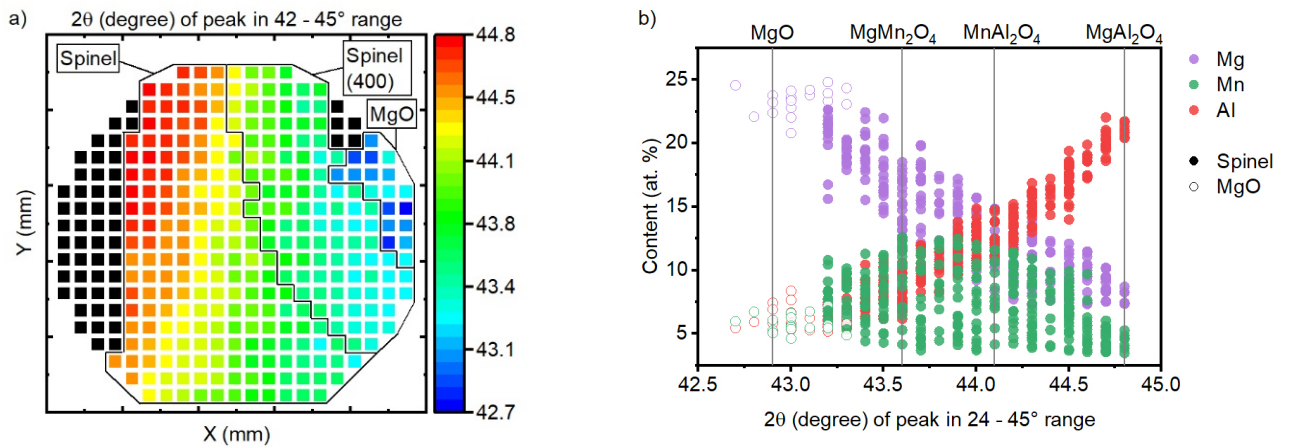

Figure 5: a) Color-coded visualization of the diffraction angle 2θ of the peak in the range of 42 – 45°. X-ray amorphous MAs are indicated by black squares. The three groups of MAs with the same crystal structures (spinel, spinel (400) and MgO) are indicated. b) Diffraction angle 2θ of the peak in the range of 42 – 45° in dependance of the Mg (purple dots), Mn (green dots) and Al (red dots) contents. Solid dots indicate spinel MAs and hollow dots indicate MgO MAs. The vertical lines indicate the reference diffraction angles of (200) MgO, and (400) of the cubic spinels $MgMn_2O_4$, $MnAl_2O_4$ and $MgAl_2O_4$.

Analyzing the peak shift in more detail, Fig. 5 a) gives an overview of the peak position in the 2θ = 42 – 45° range across all MAs (X-ray amorphous MAs are depicted as black squares). Fig. 5 b) shows the correlation of the peak shift to the elemental contents of Mg, Mn and Al. The spinel (400) peak shifts in a 2θ range of 43.2 – 44.8°, which includes the reference (400) diffraction angles of stoichiometric $MgMn_2O_4$ (43.6°), $MnAl_2O_4$ (44.1°) and $MgAl_2O_4$ (44.8°) (see Fig. 5 b). Considering the compositions however, it is unlikely that there are stoichiometric spinels in the ML, but rather spinel solid solution. Where the measured (400) spinel peak agrees with the reference diffraction angle of (400) $MgAl_2O_4$ for example, the thin film composition is $Mg_{7.4-8.7}Mn_{3.5-5.3}Al_{20.4-21.7}O_{65.9-67.8}$. Diffraction angles in agreement with the reference (400) diffraction angle of $MgMn_2O_4$ were measured for compositions of $Mg_{13.2-18.5}Mn_{4.1-12.6}Al_{6.2-11.5}O_{65.2-67.6}$. While stoichiometric $MgMn_2O_4$ forms a tetragonal spinel due to the Jahn-Teller deformation [17–20], no tetragonal spinel was found in the ML. This can be attributed to the constant presence of Al, since substitution of $Mn^{3+}$ by $Al^{3+}$ can suppress the Jahn-Teller deformation [17,19].

Fig 5 b) shows that there is no clear correlation of the Mn content to the peak shift, whereas increasing Mg contents correlate to decreasing diffraction angles, and vice versa in case of Al. Considering the atomic radius of Mg being larger than that of Al [24,25], the results correspond to Vegard's law [26]: Increasing Mg content in the spinel solid solution causes increasing lattice parameters and thus decreasing diffraction angles, and vice versa in case of Al.

The (200) MgO peak is observed to shift within 2θ = 42.7 – 43.3° (see Fig. 5 b), which is possibly caused by MgO solid solution with Mn and Al, since 4.6 – 7.2 at. % Mn and 5.2 – 8.4 at. % Al are present in the MAs identified to have MgO structure. The MgO-MnO quasi-binary phase diagram shows a wide miscibility gap (two phase region MgO + MnO) at temperatures below 500°C [21], and the limited solubility of $Al_2O_3$ in MgO is restricted to high temperatures of about 1600 – 2400°C [12,27]. However, the as-deposited ML may be in a non-equilibrium state, so the phases can differ from the equilibrium phases suggested by the phase diagrams.

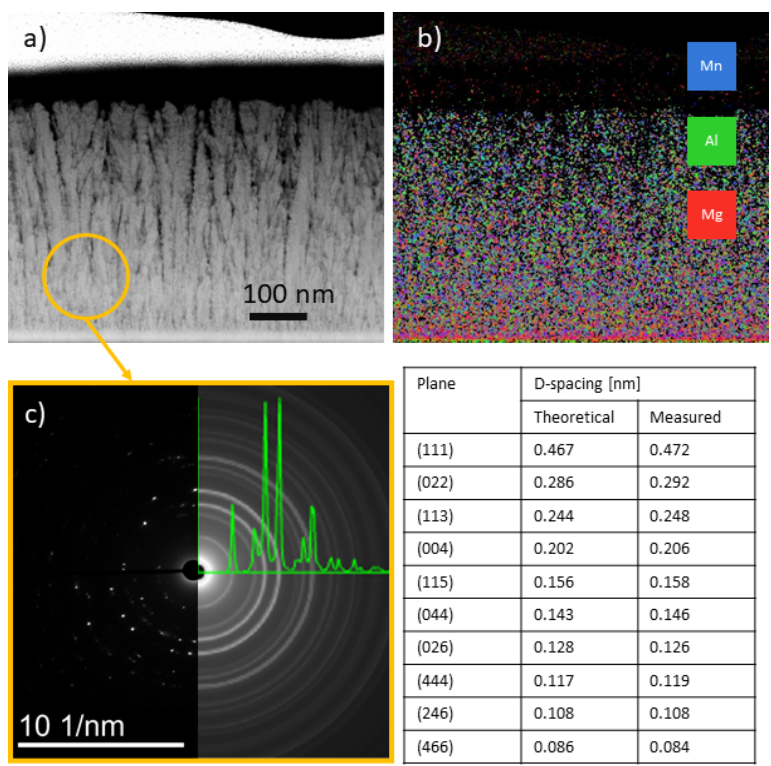

Figure 6: a) STEM DF image and b) the corresponding STEM EDX map, c) electron diffraction pattern of the circled region from a), combined with the rotational averaging pattern and the identified peaks. The table contains the theoretical $MgAl_2O_4$ spinel d-spacings (ICSD 411021) and the measured ones.

The MA selected for the TEM investigation contains equal atomic amounts of Mg, Mn and Al. The results confirm the cubic spinel solid solution. Fig. 6 a) shows a DF STEM image, 6 b) the corresponding EDX distribution map, and 6 c) the electron diffraction pattern, combined with the rotational averaged ring pattern, along with a table containing theoretical and measured d-spacings. Figs. 6 a) and b) share the same scale bar. The thickness of the film is 420 nm. The spinel grains are columnar with widths in the nm range. The diffraction pattern in c) was obtained from the circular region marked in a), containing multiple grains. The measured distances from the ring pattern agree with $MgAl_2O_4$ spinel (ICSD 411021). The small deviations from the stoichiometric spinel can be explained by formation of a solid solution, since the present spinel also contains Mn. Mg, Mn and Al are shown to be evenly distributed and no segregated grains with different composition are visible in the EDX map (Fig. 6 b).

The near-surface compositions of the 37 XPS-investigated MAs yielded composition gradients that are qualitatively similar to the EDX results (Mg max/min at right/left, Mn max/min at bottom/top, Al max/min at upper left/lower right, respectively). Quantitatively, there are differences between XPS and EDX, see Fig. 7. For comparability with the EDX results, C was disregarded in the XPS quantification. The near-surface region (XPS results) contains 2.2 – 11.5 at. % less Mg, 2.5 at. % less – 3.4 at. % more Mn, 3.2 – 7.6 at. % more Al and 4.0 at. % less – 8.1 at. % more O compared to the thin film volume (EDX results). The deviations of the Mg, Mn and Al contents determined by XPS and EDX appear linear, as shown by the good linear fits in Figs. 7 a) – c). Examples for the peak fitting are presented in Figs. 8 a) and c) – f), along with a visualization of the mean Mn oxidation state across the ML in Fig. 8 b) (MAs not measured by XPS are depicted as grey squares in Fig. 8 b). The shown example spectra belong to the X-marked MA from Fig. 8 b), that has a volume composition of $Mg_{6.1}Mn_{6.1}Al_{20.1}O_{67.7}$. The mulitplet splitting of Mn 2p3/2 was used for the chemical state analysis of Mn, applying fitting parameters from [28]. A combination of the $Mn^{2+}$ (MnO) and $Mn^{3+}$ ($Mn_2O_3$ and MnOOH) multiplets led to excellent fits for all measured MAs. The average Mn oxidation state across all MAs is between 2.3 and 2.8. The measured Al 2p peaks could all be fitted with one symmetrical component (G*L mix = 0.3, FWHM = 1.3 – 1.6) at quite consistent binding energies (BE) of 73.6 – 73.8 eV. The C 1s spectra were fitted considering the adventitious compounds C-C, C-H (charge corrected to be at BE = 284.8 eV), C-OH, C-O-C (1.5 eV above C-C, C-H), C=O (3.0 eV above C-C, C-H) and O-C=O (4.0 eV above C-C, C-H), all with G*L mix of 0.3 and FWHM constrained to be equal. Additionally, 17 MAs required one further C 1s component corresponding to metal carbonates (5.0 eV above C-C, C-H). These carbonate species coincide with high Mg contents and are therefore ascribed to $MgCO_3$ surface contamination. This is also indicated by the Mg 1s spectra, which were fitted using one component for oxidized Mg (BE = 1303.7 – 1304.3 eV, G*L mix = 0.2, FWHM = 1.6 – 1.8 eV), and another one for $MgCO_3$ (BE = 1305.2 – 1305.5 eV eV, G*L mix = 0.2, FWHM = 1.2 – 1.5 eV). The contribution of $MgCO_3$ to the total area of the Mg 1s peaks varied in a range of 0.0 – 14.0 % (3.8 % on average) and tends to increase with increasing near-surface Mg content.

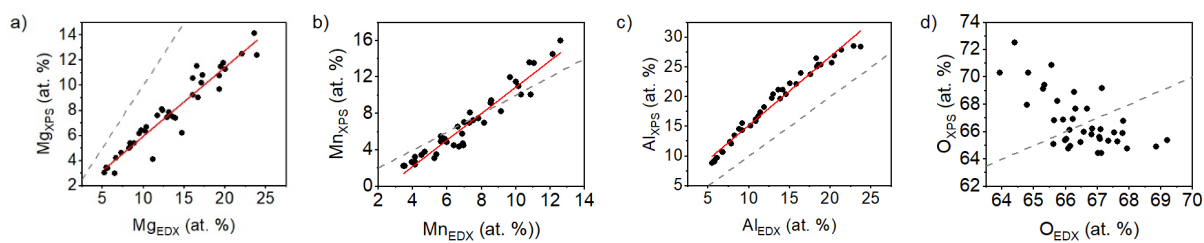

Figure 7: Correlation of the elemental contents determined by EDX and XPS of a) Mg, b) Mn, c) Al and d) O. Each dot represents one MA that was analyzed by both EDX and XPS. The dashed lines indicate equal EDX and XPS values. The red lines in a), b) and c) are linear fits of the data points.

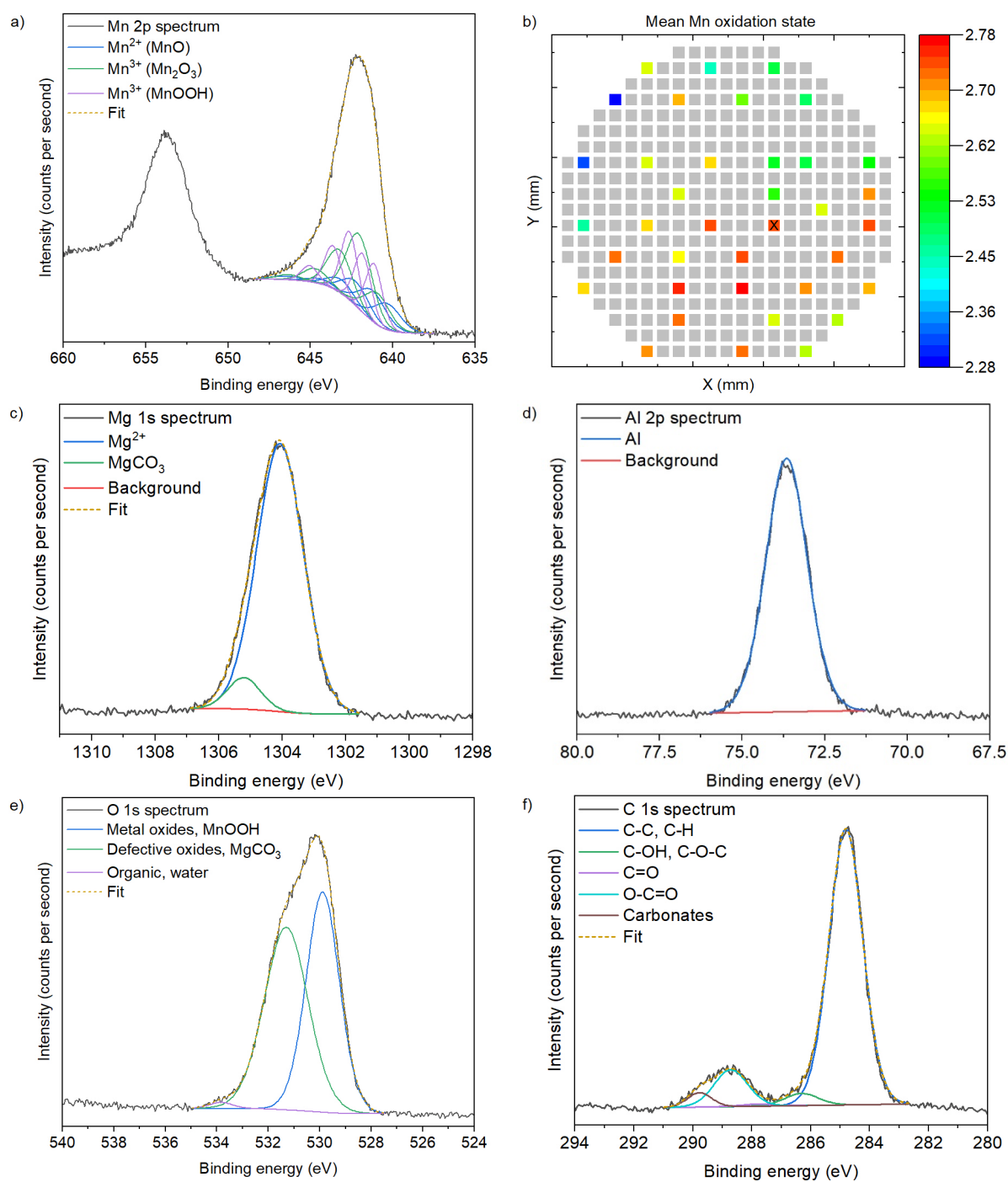

Figure 8: Example XPS spectra including peak fitting results of a) Mn 2p, c) Mg 1s, d) Al 2p, e) O 1s and f) C 1s. In b), the mean Mn oxidation state, as determined by the Mn 2p peak fitting, is visualized across the ML using a color code. All shown spectra belong to the X-marked MA from b).

The O 1s peaks were fitted using three components (all G*L mix = 0.3). The first one (BE = 529.8 – 530.1 eV, FWHM = 1.4 – 1.6 eV) is ascribed to metal oxides and MnOOH, the second one (BE = 531.0 – 531.4 eV, FWHM = 1.8 – 2.1 eV) to defective oxides and $MgCO_3$ (where present), and the third one (BE = 533.2 – 534.0 eV, FWHM = 1.1 – 1.8 eV) to surface contamination of organic compounds and possibly absorbed water.

**Conclusions**

The formation of the cubic spinel solid solution phase was identified at compositions with Mg contents of roughly 22 – 63 at. % within the tested composition range of $(Mg_{14-69}Mn_{11-38}Al_{14-74})O_X$. With increasing Mg content, preferential orientation along (400) was found. The TEM investigations showed that the approximately 420 nm-thick film contains spinel grains, which have columnar shapes and contain evenly distributed Mg, Mn and Al. At Mg contents larger than approximately 63 at %, the crystal structure changed from spinel to MgO. The Mn speciation across the whole ML was identified to be a mixture of $Mn^{2+}$ and $Mn^{3+}$. Despite the presence of Jahn-Teller-active $Mn^{3+}$, no tetragonal spinel was found.

These results show that in the typical slag subsystem Mg–Mn–Al–O, spinel solid solution is the most likely crystal structure for a wide range of the composition space. Since $Mn^{2+}$ and $Mn^{3+}$ were shown to be part of the spinel solid solution, avoiding these species in slags might enable the formation of the possible EnAM phases $Li_2MnO_3$ and/or $LiAlO_2$. However, our initial hypothesis, that $Mn^{4+}$ does not form oxides with Mg or Al, remains to be addressed in further studies, as no such species were found in this ML. This will be targeted next, possibly by applying a $Mn^{4+}O_2$ sputter target and appropriately high $O_2$ partial pressure to fabricate a ML of similar composition space as in this study but containing $Mn^{4+}$ instead of $Mn^{2+/3+}$.

**Declaration of competing interest**

The authors declare that they have no known competing financial interests or personal relationships that could have appeared to influence the work reported in this paper.

**Acknowledgements**

This research was funded by DFG SPP 2315, and project number 470309740: LU 1175/36-1, SCHI 1522/1-1. The Center for Interface Dominated High-Performance Materials (ZGH, Ruhr University Bochum) is acknowledged for the access to the XRD, FIB and TEM.